\documentclass[sn-basic]{sn-jnl}% Basic Springer Nature Reference Style/Chemistry Reference Style
\usepackage{graphicx}%
\usepackage{multirow}%
\usepackage{amsmath,amssymb,amsfonts}%
\usepackage{amsthm}%
\usepackage{mathrsfs}%
\usepackage[title]{appendix}%
\usepackage{xcolor}%
\usepackage{textcomp}%
\usepackage{manyfoot}%
\usepackage{booktabs}%
\usepackage{algorithm}%
\usepackage{algorithmicx}%
\usepackage{algpseudocode}%
\usepackage{listings}%

\begin{document}

\title[Article Title]{Titanium Oxide Absorption as a Proxy to Detect Long term Variation and Activity Cycle in Proxima Centauri
}


\author*[1]{\fnm{Fatemeh} \sur{Azizi}}\email{f.azizi@pnu.ac.ir}

\author[2]{\fnm{Mohammad Taghi} \sur{Mirtorabi}}\email{torabi@alzahra.ac.ir}

\author[1]{\fnm{Rahimeh} \sur{Foroughi}}\email{r.froghi@gmail.com}

\affil*[1]{\orgdiv{Department of Physics}, \orgname{Payame Noor University}, \orgaddress{\city{Tehran}, \country{Iran}}}
%\affil*[1]{\orgdiv{Department of Physics}, \orgname{Payame Noor University}, \orgaddress{\city{Tehran}, \postcode{}, \country{Iran}}}
\affil[2]{\orgdiv{Department of Fundamental Physics, Faculty of Physics}, \orgname{Alzahra University}, \orgaddress{\city{Tehran}, \country{Iran}}}

%\affil[3]{\orgdiv{Department}, \orgname{Organization}, \orgaddress{\street{Street}, \city{City}, \postcode{610101}, \state{State}, \country{Country}}}


\abstract{Stellar activity cycles on magnetically active stars can be estimated by molecular absorption bands. We have previously introduced a molecular index which compares absorptional line strength of the $TiO\lambda567nm$ with its nearby continuum has previously been introduced. In this work we use this indicator to evaluation long-term activity variations for Proxima Centauri star, using spectroscopic data from HARPS. The results indicate periodicity with an activity period of $2873_{-53.9}^{+47.4}$ days, which is similar to the previous measurements from other indicators. 
}

%%================================%%
%% Sample for structured abstract %%
%%================================%%




\maketitle

\section{Introduction}\label{sec1}

Stellar activity cycles are commonly observed in both the Sun and many late-type stars. These cycles are believed to be driven by magnetic activity, which can reflect various parameters of the star such as magnetic field intensity, interior structure, and rotation. Investigating these cycles can provide essential details on the dynamo mechanism, although many aspects of this process remain poorly understood. Stellar activity is also closely linked to the exploration of exoplanets since star spots can mimic or even obscure the detection of planets \citep{queloz2001}. With modern spectrographs, radial velocity measurements can now reach precision of one meter per second \citep{mayor2003} or even more precisely  ten centimeter per second \citep{pepe2013}. However, at such precision levels, signals in the radial velocity curves can be caused by the stellar activity and become  a significant limiting factor in the search and further studies of Earth-like exoplanets.

Since 1966, the HK Project has been conducting research at Mount Wilson Observatory (MWO) using chromospheric $CaII\ H\&K$ lines to monitor the intensity and extent of stellar magnetic fields \citep{wilson1978,balinus1995}. The project aims to search for indications of cyclic magnetic activity and is known for being the longest-running plan for this purpose. To find analogs of the solar 11-year activity cycle, the project has optically monitored hundreds of stars. However, due to their optical faintness, particularly in the $CaII\ H\&K$ lines, which have been the foundation of most studies, almost no M-type star has been examined. The lack of results for M-type stars is not due to a lack of interest, but rather the difficulty in finding appropriate criteria for investigating possible signs of activity.

Cool stellar atmospheres, especially in late M stars, are characterized by molecular bands. A diverse range of molecular lines can be observed in the spectra of sunspots and cool stars \citep{wallace2000}, including diatomic molecules lines, which can indicate temperature and pressure. These lines are also useful for determining elemental abundances. The magnetic properties of molecular band systems in visible and near-infrared spectra of sunspots and cool stars have been examined by \citet{berdyugina}, and it has been shown that several molecular lines, such as Titanium oxide (TiO), can serve as suitable indicators of the stellar magnetic fields.
TiO is a molecule that is temperature-sensitive and forms in atmospheres with temperatures cooler than 4000 K. It can be used as an indicator of spots in hotter atmospheres. TiO absorption is the strongest feature in the near-infrared region of stellar spectra, and it can be observed even with low-precision spectrometers or filtered photometers. According to \citet{mirtorabi2003}, TiO absorption was observed in the atmosphere of variable star $\lambda$ Andromeda, and an activity cycle of 4-14 years was estimated. Additionally, TiO absorption remains present even when the star is at its brightest, suggesting that active areas may be distributed evenly on the surface of the star.

As a proxy to display spots and activity features, a photometric system based on one of the major absorption of $TiO$ molecules in 719 nm was introduced by \citet{win1992}. This photometric system has recently been updated by \citet{azizi2015} by a slight shift in continuum bands of Wing system in order to make it less contaminated from the weak absorption of $VO$ bands. In our previous paper \citep{bidaran2016} a new TiO index was introduced which examines the absorption at visual bands around 567 nm and then applied it to investigation for stellar activity in sample of solar-type stars \citep{azizi2018}. In this work, we attempted to determine possible long-term activity cycles of the  closest star to the Sun,  $Proxima \ Centauri$. For this purpose, we used HARPS high-resolution spectra spanning 13 years. In section 2 a brief history of proxima centauri observation is given. The specification of the $B$\_~index and the detail of data and analysis are presented in Section 3.  Finally, we discuss and summarize our results in Section 4.

%The Introduction section, of referenced text \cite{bib1} expands on the background of the work (some overlap with the Abstract is acceptable). The introduction should not include subheadings.

%Springer Nature does not impose a strict layout as standard however authors are advised to check the individual requirements for the journal they are planning to submit to as there may be journal-level preferences. When preparing your text please also be aware that some stylistic choices are not supported in full text XML (publication version), including coloured font. These will not be replicated in the typeset article if it is accepted. 

\section{Proxima Centauri}

Proxima Centauri, also referred to as $V645 \ Cen = GJ~551 = \alpha \ Cen~C$, is a nearby $M5.5V$ star with a magnitude of 11.3 in the V-band \citep{benedict1993}. It is located at a distance of 1.305 pc \citep{luric2014}, making it the closest star to the Sun and easily observable. This star has been regularly monitored in the optical and UV bands and has been tentatively reported as a variable star with periods ranging from 1.2 to 7 years. The $H_\alpha$ line of this star was measured over a period of seven years using observations made by the 2.15m telescope of CASLEO and reported a potential 1.2  years cycle \citep{cincunegui2007}. A 3-year cycle was proposed by analyzing five years of data from the Hubble Fine Guidance Sensors, \citep{benedict1998}.
%\citep{benedict1998}
%A possible activity cycle in Proxima Centauri has been identified with a 1.2 year cycle period and a false alarm probability of 35\% by \citet {sturrock2010}.

 An optical evidence for an activity cycle duration of $6.9\pm 0.5$ years has been seen by \citet{jason2007} based on implementing five years of $V$-band data from ASAS-3,  observed as a part of  All Sky Automated Survey \citep{pojmanski2002},  and combined with ultraviolet data from the International Ultraviolet Explorer (IUE) and the Far Ultraviolet Spectroscopic Explorer (FUSE) missions. \citet{guinan2010} has updated this period to be 7.6 years later. A wide peak around 8 years was determined by further utilization of power spectrum of 9 years data of ASAS by \citet{savanov2012}. 
They also measured an amplitude power spectra, based on measurements of the $CaII \ H\&K$ lines. \citet{suarez2016} analyzed ASAS-3 data and discovered activity cycles in Proxima Centauri with period of $6.8\pm 0.3$ years. 
\citet{wargelin2017} analyzed several years of optical, UV, and X-ray observations of Proxima Centauri. They find evidence for a 7 years stellar activity cycle and an 83 days rotational period for this star.

A research conducted by \citet{gomes2011,gomes2012}  analyzed 29 $M0-4$ stars, including Proxima Centauri, using data collected by HARPS spectrometer over seven years. The study discovered that almost one-third of the stars (excluding Proxima Centauri) showed long-term variability in certain lines such as $CaII$, $H_\alpha$, $HeI\  D3$, and $NaI \ D$. \citet{anglade2016} used HARPS along with spectral and photometric data to study Proxima Centauri, and found no cycle but observed an approximately 80-day rotational periodicity. \citet{collins2017} investigated the possibility of recognizing periodicity in the $H_\alpha$ line in high-resolution spectra, such as those derived from HARPS. They found the rotation period, but no long-term cycle in $H_\alpha$. $H_\alpha$ line's behavior is a potentially significant diagnostic because it is sensitive to magnetic activity and is a strong emission line typically observed in the spectrum of late M-dwarf stars.

\section{Method and Analysis}

Potential evidence of active regions on solar-type stars has been found through studies using molecular bands of TiO \citep{azizi2018, mirtorabi2003}. The aim of our research is to identify significant long-term periodicities in the light curve of Proxima Centauri by analyzing the variation in absorption of $TiO$ at a wavelength of 567 nm. We examined 170 spectra collected by HARPS between February 25, 2004 and August 14, 2017, and utilized the $B$\_~index to identify any indications of periodic activity in the time series.

\subsection{B\_~Index}

\citet{bidaran2016} proposed an index that is based on the absorption of titanium oxide ($TiO$) in the visual region. The level of absorption is measured by defining two band passes, one at the core of the $TiO$ absorption at $ 567 \ nm$ with a width of $16 \ nm$ and the other on the nearby continuum at $610 \ nm$ with a width of $10 \ nm$, referred to as filter $D$ and $E$, respectively. Based on the accumulated flux in these two filters they presented an index  named $B$-index which measures the difference between magnitude in filter $D$ and $E$. $B$-index is mimicking the Wing $TiO$ index \citep{win1992}  on the visual region and defined as 
\begin{equation}
\label{eqn:1}
B \_ ~index=-2.5 ~log~ \frac{\int_{}^{}F_{D}(\lambda)~ S_{D}(\lambda)~d\lambda }{\int_{}^{}F_{E}(\lambda) ~S_{E}(\lambda)~d\lambda}
\end{equation}
where $F_D{(\lambda)}$ is the flux in the core of the line (567 nm), and $F_E{(\lambda)}$ is the flux on the nearly continuum (610 nm). $S_D{(\lambda)}$ and $S_E{(\lambda)}$ are appropriate filter response function. To quantify the $B$-Index as an marker of cool regions on the surfaces of active stars, a $B$-index versus $T_{eff}$ calibration was performed by \citet{bidaran2016}, using Kitt Peak National Observatory (KPNO) spectra for cool stars. 
According to Figure 4 in their  paper, the $B$\_~index levels of stars decrease as their temperature increases. When the temperature of the atmosphere goes beyond $4000  \ K$, the TiO molecules become disassociated, leading to the absence or weakness of the absorption feature, causing the $B$\_~index to approach zero. Conversely, the $B$\_~index increases as the temperature drops, eventually reaching a value of around 2, which corresponds to a spectral type of M7.5.

\subsection{Spectroscopic data}

We looked through the public data archive of HARPS ESO to find available spectra of TiO molecules at the 567 nm absorption line for Proxima Centauri. We chose all observed spectra from February 2004 to August 2017, which resulted in 170 individual spectra, to calculate our $B$-index. These observations cover a time span of over 13 years, providing a sufficient timeframe to indicate an activity cycle for this star. Table \ref{tbl:01} displays the Julian date of observation and the calculated $B$-index for the selected spectra.

\begin{figure}
%\vspace{15pt}
\centering
\includegraphics[width=0.9\textwidth]{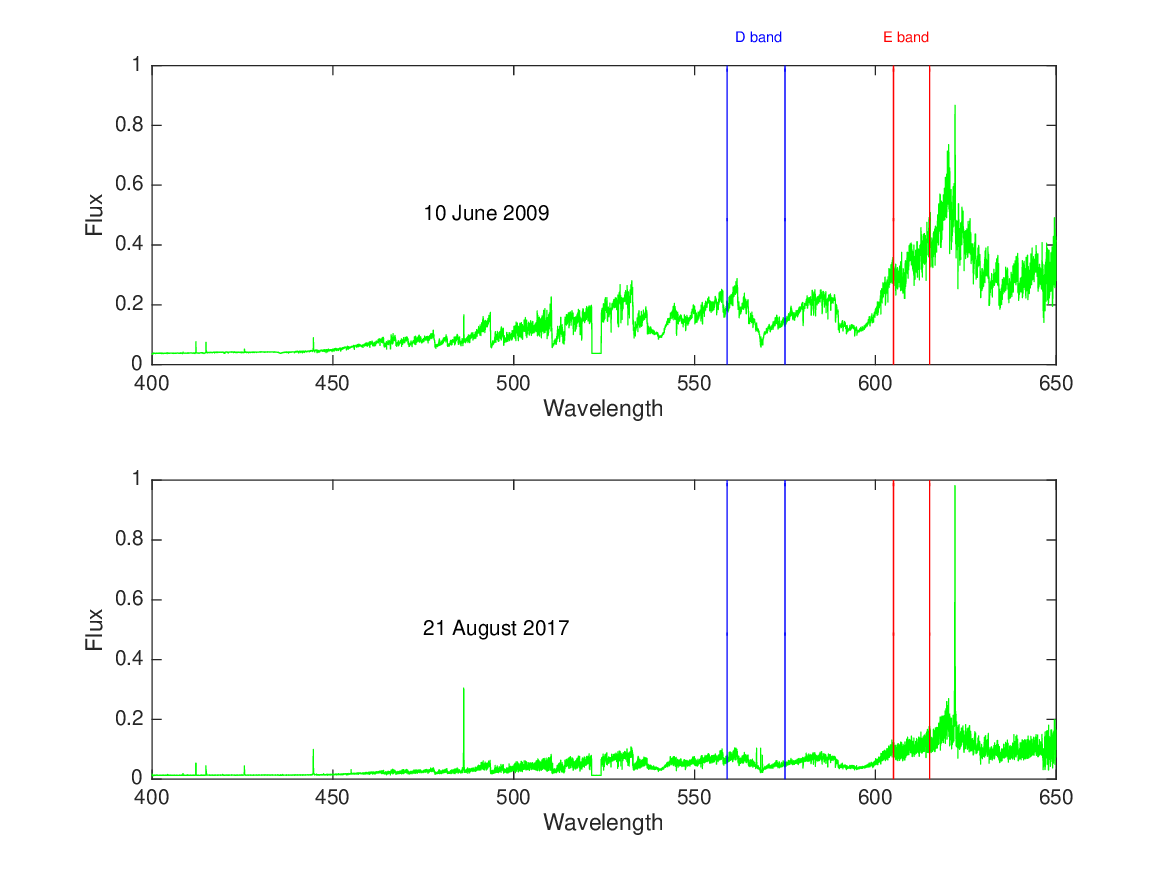}
\caption{$TiO$ absorption variation during 8 years on surface of proxima centauri. Vertical lines depicts regions were used to measure $B$\_~index.}
\label{fig:01}

\end{figure}

The HARPS is a high-resolution spectrograph that employs fiber-fed technology to achieve high accuracy radial velocity measurements. It is situated at the 3.6m ESO telescope in LA Silla observatory, Chile \citep{mayor2003}. Its resolving power is about $R\sim 115,000$, and it covers a spectral range of 378 to 691 nm. We used the wavelength-calibrated spectra extracted order-by-order by the HARPS pipeline for our analysis.
%Figure~\ref{fig:01} presents $TiO$ absorption at $567 \ nm$ of a sample spectrum of Proxima Centauri taken in 15 April 2010 09:16:35 UTC and 6 May 2013 10:06:59. $D$ and $E$ band passes are indicated by vertical lines.
 In figure~\ref{fig:01}, two example spectra of the star are shown, which are taken in 2009 and 2017, evidently showing the changes in the amplitude and shape of the $TiO$ absorption. $D$ and $E$ band passes are indicated by vertical lines.

%\begin{figure}

%\centering
%\includegraphics[width=0.9\textwidth]{1.eps}
%\caption{$TiO$ absorption in spectrum of proxima centauri. Vertical lines depicts regions were used to measure $B$\_~index.}
%\label{fig:01}
%\end{figure} 

\

\subsection{Periodicity }
\label{sec3.3}
We have obtained a periodicity diagram or periodogram for this star based on the results taken from  calculations of time ordered spectra. Periodogram analysis \citep{lomb1976,scargle1982} is the most commonly used method of looking for periodicity in time ordered and non-uniformly spaced data.
It is similar to least-squares fitting of sin waves, in the form of $y=A~Cos \omega t+B~Sin \omega t$. 
The normalized periodogram analysis gives the best sinusoidal that fits the non-uniformly sampled data, and the validity of the period is deduced from it's FAP \citep{sturrock2010}. 

To analyze activity cycles, a generalized Lomb-Scargle periodogram (GLS) was computed for Proxima Centauri. This method extends the Lomb-Scargle periodogram by taking into account the amount of errors and is more suitable for time series with a non-zero mean. The GLS utilizes the formula $y=A~Cos \omega t+B~Sin \omega t + C$ and implements a power spectrum \citep{zechmeister2009}.
To assess the reliability of a periodogram peak obtained from GLS, we consider it significant if it exceeds 1\% of the FAP. This means that the peak has a 99\% confidence level that it is not caused by Gaussian noise. FAP levels are calculated by randomly permuting the data with the same number of observations. We fit activity cycles by using a simple sine wave at the period of maximum power obtained from the GLS.

\begin{figure}
%\vspace{15pt}
\centering
\includegraphics[width=0.9\textwidth]{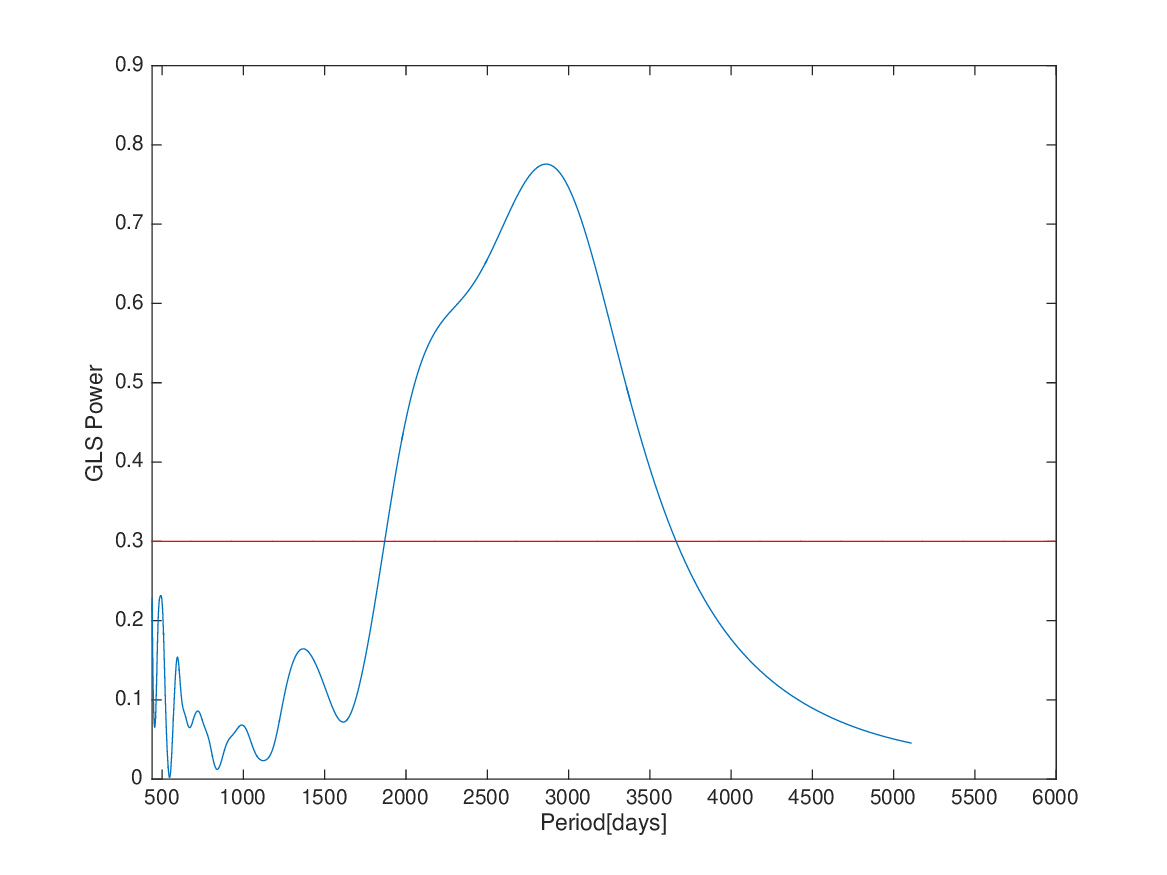}
\caption{The GLS Periodogram of $B$\_~index derived from the HARPS data for Proxima Centauri. The red line denotes the 1\% false-alarm probability.} 
\label{fig:02}
\end{figure}

\begin{figure}
%\vspace{15pt}
\centering
\includegraphics[width=0.9\textwidth]{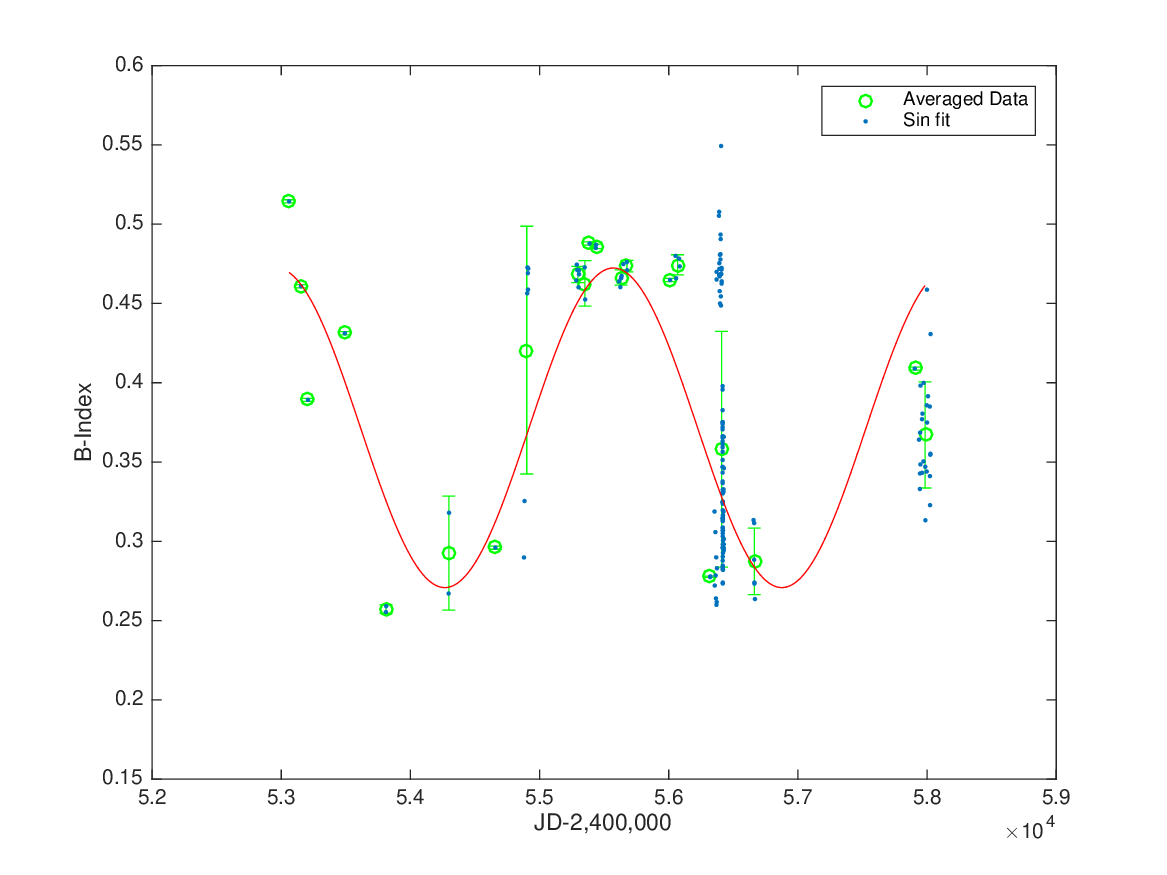}
\caption{Time series of the $B$\_~index data with the fitted activity cycle for Proxima Centauri. The blue dots are the data points and the red curve is the fit.}
\label{fig:03}
\end{figure}

The GLS periodogram shown in Figure~\ref{fig:02} displays the visual band data from 170 HARPS observations of Proxima Centauri. The red line indicates the FAP level for the strongest peaks, among which the one at approximately 2873 days is indicative of a definite stellar cycle. Figure~\ref{fig:03} shows a long-term spectroscopic cycle fit for the star with a period of approximately 2873 days and a semi-amplitude of about 0.13 mag. Yearly averages with error bars are also plotted to demonstrate the cyclic behavior, with blue dots representing individual measurements and green circles indicating the average measurements of every phase bin.

Our focus is on the activity cycle that occurs within a period range of 1.5-20 years. We have chosen 1.5 years as the lower limit for long-term periodicities, as \citet{anglade2016} suggested a period of 80 days for the rotation of this star, our lower limit of 1.5 years for the activity period is well above the rotational period and avoids any conflict.
This limit ensures that any indication of rotational mutability is appropriately removed from the final results. Since the observation time span was limited to approximately 13 years the upper limit for periodicities was set at 20 years. We used the Monte Carlo simulation to calculate certitude distances for any given period. To generate various noise realizations for each data point, we utilized a Gaussian distribution with the same standard deviation as the calculated short-term scatter. By recalculating the GLS periodograms, we obtained a certitude distance from the distribution of the derived periods.

\section{Results and Discussion}

For many decades, researchers have conducted extensive studies on the activity of stars categorized as type $F$, $G$, and $K$. However, comparatively less attention has been given to $M$ dwarfs, and only a few long-term activity cycles have been reported in existing literature, according to sources such as \citet{fuh2023}, \citet{robertson2013} and \citet{gomes2012}. 
In this study, we analyzed 13 years of HARPS data on Proxima Centauri and found major indication of a period of $2873_{-53.9}^{+47.4}$ days (approximately 7.9 years) variations, which confirms previous measurements of the period by \citet{suarez2016}, \citet{savanov2012}, and \citet{kiraga2007}. Our findings do not support the 1.2-year or 3-year periods presented by \citet{cincunegui2007} or \citet{benedict1998}, respectively. However, our 7.9-year period is in accordance with the 7 years stellar activity cycle cited by \citet{wargelin2017} and with the $6.8 \pm 0.3$ year period obtained by \citet{suarez2016} from the ASAS data set.
%\citet{endle2008}

\citet{wargelin2017} have conducted a periodicity analysis of the optical, UV, and X-ray observations of this star. They used 15 years of ASAS-3 and ASAS-4 photometry in the V band, as well as Swift XRT and UVOT observations, to confirm the previously detected 83-day rotational periodicity and also found "strong evidence" for a $2576 \pm 52$  days (~7-year) period. Comparing our results with \citet{wargelin2017}, we found that our maxima and minima are approximately in phase with theirs, but with a slightly longer period.

\citet{mirtorabi2003} have observed that the TiO light curve of the star $\lambda$ Andromedae has a longer period compared to the visual light curve. This might be due to the spatial distribution of bright and dark spots on the surface of the star. The visual light curve is more sensitive to the brighter regions,  whereas the TiO absorption features might be more prevalent at dark spots. Comparing our study with \citet{wargelin2017}  extends this phase shift and period discrepancy to the long term activity variation  where TiO absorption seems to have longer activity cycle than visual light.    

 Table \ref{tbl:01} displays the measured $B$\_~index for all 170 spectra used in the activity calculation. Figure \ref{fig:02} illustrates the calculated periodogram using the Lomb Scargle method. The peak at 2873 days is significantly above the FAP line and demonstrates a clear periodicity in the data. Figure \ref{fig:03} presents a plot of the averaged $B$\_~index points throughout the entire period, which shows significant variation in the TiO absorption during activity cycle. A sine wave with the appropriate frequency from the periodogram is also overlaid.

 \citet{suarez2016} conducted a study on the activity cycle of 22 early M-type stars and found that the average length of the cycle is 7.4 years. They also found that 9 mid M-types have a mean cycle length of 7.5 years. Our own observations of Proxima Centauri, which is a mid M-type star, are consistent with these findings. It is worth mentioning that the $CaII\ H\&\ K$ lines, which are commonly observed in $F$ to $K$ stars, are not very visible in M-stars. These lines have been used as a major indicator of activity in previous studies. As TiO is a major spectral manifestation of cool $M$-type stars, the $B$\_~index can be used as an effective indicator of activity in the absence of the usual indicators for magnetic cycles.
%and $H_\alpha$ 
\section*{Acknowledgments}

Based on observations collected with the {\it HARPS \/}  spectrograph on the 3.6-m telescope at {\it La Silla \/} Observatory,  European Southern Observatory, Chile, Science Archive Facility.
%under programme ID 072.C-0488(E). 

\section*{Data Availability}

The data underlying this article are available in  the ESO Science Archive Facility in "https://archive.eso.org/scienceportal/home".

\begin{table}[h]
\caption{The Results; Time Series of the $B$\_~index data for Proxima Centauri Star. }\label{tbl:01}
%\begin{tabular*}{\textwidth}{@{\extracolsep\fill}lccccccc}
\begin{tabular}{@{}llllllll@{}}
\toprule%
$HJD(days)$ & $B\_~Index$  & $HJD(days)$ &$B\_~Index$&$HJD(days)$& $B\_~Index$&$HJD(days)$& $B\_~Index$\\ 
\midrule
53060.36&0.5144&56361.33&0.3058&56417.23&0.375&56424.14&0.319  \\
53153.14&0.4609&56363.27&0.2785&56417.24&0.366&56425.99&0.0.298 \\   
53207.00&0.3892&56365.31&0.264&56417.27&0.357&56426.035&0.296\\
53492.20&0.4312&56367.27&0.2899&56417.30&0.35 &56426.08&0.302\\
53810.28&0.2554&56369.25&0.2599 &56417.35&0.383&56426.101&0.294\\
53812.28&0.2591&56370.22&0.4699&56417.370&0.337&56426.23&0.366\\
54296.09&0.2671&56370.38&0.4651&56417.392&0.398&56426.97&0.346\\
54299.09&0.318&56371.30&0.2617&56417.399&0.347&56656.31&0.313\\
54658.06&0.296&56373.23&0.283&56417.998&0.283&56661.30&0.288\\
54879.32&0.2898&56388.35&0.5053&56418.01&0.297&56661.35&0.312\\
54883.36&0.3254&56390.18&0.468&56418.018&0.299&56663.35&0.274\\
54904.27&0.4564&56390.36&0.5078&56418.024&0.325&56663.37&0.273\\
54906.32&0.4727&56391.25&0.472&56418.03&0.303&56667.28&0.264\\
54909.32&0.4691&56391.39&0.475&56418.055&0.309&57905.02&0.409\\
54910.32&0.4587&56394.21&0.467&56418.08&0.325&57937.13&0.364\\
54912.33&0.472&56394.38&0.458&56418.09&0.307&57944.04&0.333\\
55284.27&0.4645&56396.33&0.481&56418.10&0.288&57945.08&0.343\\
55288.29&0.4745&56397.12&0.45&56418.12&0.296&57946.03&0.369\\
55291.34&0.4711&56400.13&0.478&56418.14&0.330&57948.14&0.349\\
55301.31&0.4602&56400.31&0.48&56418.15&0.314&57949.16&0.399\\
55304.27&0.4709&56401.103&0.493&56418.16&0.320&57960.98&0.377\\
55305.24&0.4684&56402.12&0.491&56418.17&0.324&57961.96&0.343\\
55349.08&0.4728&56402.35&0.4686&56418.19&0.325&57963.96&0.381\\
55352.94&0.4525&56403.1&0.455&56418.205&0.333&57971.96&0.35\\
55387.06&0.4878&56404.099&0.449&56418.25&0.305&57972.96&0.4\\
55434.10&0.485&56405.34&0.549&56418.26&0.314&57986.01&0.347\\
55436.10&0.487&56407.23&0.469&56418.27&0.315&57987.02&0.313\\
55614.37&0.4639&56408.273&0.471&56418.29&0.313&57996.96&0.344\\
55625.33&0.4603&56409.085&0.463&56418.304&0.308&57997.98&0.386\\
55630.32&0.4659&56409.280&0.472&56418.31&0.290&57998.97&0.459\\
55631.33&0.4674&56410.082&0.464&56418.32&0.282&57999.97&0.375\\
55635.35&0.2776&56417.004&0.347&56418.34&0.274&58007.02&0.392\\
55648.35&0.4748&56417.033&0.372&56418.36&0.285&58020.958&0.385\\
55675.20&0.476&56417.041&0.365&56418.37&0.284&58021.96&0.341\\
55678.17&0.471&56417.055&0.371&56419.161&0.315&58022.96&0.323\\
56010.24& 0.4648&56417.078&0.396&56419.167&0.316&58024.968&0.355\\
56052.25&0.48&56417.086&0.375&56420.048&0.282&58025.974&0.355\\
56057.29&0.4659&56417.093&0.363&56420.054&0.307&58026.98&0.431\\
56079.11&0.4783&56417.12&0.361&56420.062&0.293& &\\
56085.17&0.4734&56417.130&0.374&56420.093&0.301&&\\
56320.36&0.2774&56417.162& 0.360&56420.099&0.316&&\\
56322.32&0.2779&56417.17&0.343&56420.15&0.295&&\\
56354.31&0.3188&56417.184&0.338&56423.98&0.333&&\\
56356.30&0.2721&56417.19&0.331&56423.99&0.331&&\\
\midrule
Period of & & & & & & & \\
$B$\_~Index (days)& =$2873_{-53.9}^{+47.4}$& & & & & &\\
\botrule
HJD=JD-2,400,000\\
%\end{tabular*}
\end{tabular}
\end{table}

\end{document}